\documentclass[aps,prd,onecolumn,showpacs,superscriptaddress,nofootinbib]{revtex4-1}

\usepackage[pdftex]{graphicx}

\usepackage{amsmath}
\usepackage[latin1]{inputenc}
\usepackage{txfonts}
\usepackage{hyperref}

\usepackage{xcolor}

\newcommand{\dottedcolumn}[3]{%
	\settowidth{\dimen0}{$#1$}
	\settowidth{\dimen2}{$#2$}
	\ifdim\dimen2>\dimen0 \dimen0=\dimen2 \fi
	\begin{pmatrix}\,
		\vcenter{
			\kern.6ex
			\vbox to \dimexpr#1\normalbaselineskip-1.2ex{
				\hbox{$#2$}
				\kern3pt
				\xleaders\vbox{\hbox to \dimen0{\hss.\hss}\vskip4pt}\vfill
				\kern1pt
				\hbox{$#3$}
			}\kern.6ex}\,
	\end{pmatrix}
}

\begin{document}

\title{Remote synchronization in networks of coupled oscillators}
\author{Rafael S. Pinto}
\email{rsoaresp@gmail.com}
\affiliation{Institute of Mathematical and Computer Sciences, Universidade de São Paulo, São Carlos 13566-590, São Paulo, Brazil}

\date{\today}

\begin{abstract}
We study under which conditions systems of coupled oscillators on complex networks display \emph{remote synchronization}, a situation where pairs of vertices, not necessarily physically linked, but with the same network symmetry, are synchronized.
\end{abstract}

\maketitle

\section{Introduction}

How the patterns in the structure of naturally occurring systems affect the dynamics that they are supposed to perform is one of the most active areas of research nowadays, with the study of how synchronous behavior may emerge in systems formed by heterogeneous elements taking a prominent role \cite{pikovsky2001,winfree1980}. Some the applications range from the small scales of metabolic processes in populations of yeast \cite{monte2007}, coupled flagella \cite{wan2014} and brain dynamics \cite{dotson2016} to the large scale of power grids \cite{motter2013}.

Recently, two papers appeared that highlighted how symmetries in the pattern of connection in networks of oscillators reflect on the dynamics. Consider a star, the network depicted on the left in Figure \ref{sym_examples}. It has a central vertex, know as \emph{hub}, that has $n$ neighbors, the \emph{leaves}, that have no other connection than with the hub. On top of a star, the Stuart-Landau model of coupled oscillators is neatly given in complex form by

\begin{equation}
\begin{split}
\dot{z}_h = (\alpha+ i\omega_h - |z_h|^{2})z_h + \lambda \sum_{j=1}^{n} (z_j - z_h),\\
\dot{z}_i = (\alpha+ i\omega_i - |z_i|^{2})z_i + \lambda  (z_h - z_i),
\end{split}
\label{sl_complex_model}
\end{equation}  
where $z_k = x_k + iy_k$ is the (complex) state of the $k$th oscillator (with index $h$ standing for the hub), $\omega_k$ is the corresponding natural frequencies and $\lambda$ the coupling strength. This model was analyzed in \cite{bergner2012}, where it was found, both numerically and experimentally, that if the leaves have approximately the same natural frequency and the hub has a sufficiently large frequency mismatch, a regime called \emph{remote synchronization} (RS) appears, a situation where the leaves can synchronize their oscillations even if the hub does not (the same color of the leaves represent their common rhythm in Figure \ref{sym_examples}). 

This scenery, where units not physically connected can behave in unison, even if intermediaries elements are out of synchrony, was latter also seen in some complex networks of SL oscillators \cite{gambuzza2013}. 

Almost at the same time a similar phenomenon was observed for the Kuramoto-Sakaguchi model \cite{sakaguchi1986,kuramoto1975,strogatz2000},

\begin{equation}
\dot{\varphi}_i = \omega_i + \lambda \sum_{j=1}^{N} A_{ij} \sin(\varphi_j - \varphi_i - \alpha),
\label{sakaguchi_kuramoto model}
\end{equation}
for general networks described by the adjacency matrix with elements $A_{ij}$ and in the case of identical oscillators (same natural frequency, $\omega_i = \omega$, for all oscillators). The results found in \cite{nicosia2013} were that elements of the network not directly linked could display the same phase, even if intermediaries vertices would lock in different ones. An example of this behavior is shown for the network on the right in Figure \ref{sym_examples}, where colors represent phases. An argument was elaborated in \cite{nicosia2013} of why this happens for the linearized version of the Kuramoto-Sakaguchi model, involving the commutation relations between permutation matrices representing automorphisms of the graph and the corresponding Laplacian matrix. 

The role of the graph automorphisms was further elaborated for identical coupled oscillators in \cite{pecora2014}, but the question of which conditions a general model, possibly with heterogeneous parameters, must satisfy to present remote synchronization is still far from obvious

Furthermore, \cite{vlasov2016} also found RS to occur in variants of the Kuramoto-Sakaguchi on real world networks with degree-frequency correlation, a condition necessary to emulate the frequency mismatch used in \cite{bergner2012} for the star graphs.

The picture for RS is unclear at the moment. Are the results presented in \cite{bergner2012,gambuzza2013,nicosia2013,pecora2014,vlasov2016} originated by the same mechanism? For Stuart-Landau oscillators on a star, the free amplitude of the hub was pointed in \cite{bergner2012} as the key to the emergence of RS, as it could transmit information between leaves, allowing their synchronization. This argument, however, would forbid RS in phase oscillators. On the other hand, whereas RS was found for identical phase oscillators in \cite{nicosia2013,pecora2014}, in \cite{vlasov2016} the natural frequency distribution may be quite heterogeneous due to the degree frequency correlation imposed and the fact that it is not unusual to find real world networks with the presence of massively connected vertices. 

In this paper we show that the mentioned previous results are in fact originated by the same mechanism and moreover the idea originated in \cite{nicosia2013,pecora2014} is correct, the results observed  in the cited references are an artifact of the graph symmetries, described by its automorphisms. Our calculations, it is important to stress, are valid not only in the linear regime, and can be extended to a wide class of phase oscillators models, as well as networks of coupled Stuart-Landau oscillators, covering many important examples in the literature. In the case of an heterogeneous ensemble of oscillators, we can show which conditions must be satisfied by these parameters in order to observe RS.

Our conclusions were achieved by first rewriting the equations of motion in a matrix-vector notation that is more suitable to be analyzed under permutations matrices $P$ that are graph automorphisms. Once we know how $P$ affects the equations of motion, the permutations can be used to partition the vertices into disjoint sets such that all the elements in each one of these sets must be governed by the same dynamics, originating remote synchronization.

This paper is organized as follows. In section (\ref{preliminaries}) we will discuss some essential facts that will be necessary to obtain our results. Remote synchronization for a  large class of phase-oscillators models is studied in (\ref{remote_sync_phase_oscillators}) and section (\ref{remote_sync_sl_oscillators}) extends this result to the Stuart-Landau model. Some numerical examples are shown in section (\ref{examples}) with our conclusions presented in (\ref{conc}).

\section{Preliminaries}
\label{preliminaries}

In all of our calculation we will consider only undirected and unweighted networks $\mathcal{G} = (V, E)$ with $n = |V|$ vertices, $m = |E|$ links and adjacency matrix $A$ with entries $A_{ij} = 1$ if vertices $i$ and $j$ are connected and $A_{ij} = 0$ otherwise. The degree $k_i$ of vertex $i$, its number of neighbors, is given by $k_i = \sum_{j=1}^{n} A_{ij}$. By graph symmetries we mean \emph{graph automorphisms}, bijections from the set of vertices to itself such that it keeps the structure of $A$, it preserves the vertex-link connectivity in such a way that the adjacency matrix is the same. 

\begin{figure}
	\centering
	\includegraphics[width=0.7\linewidth]{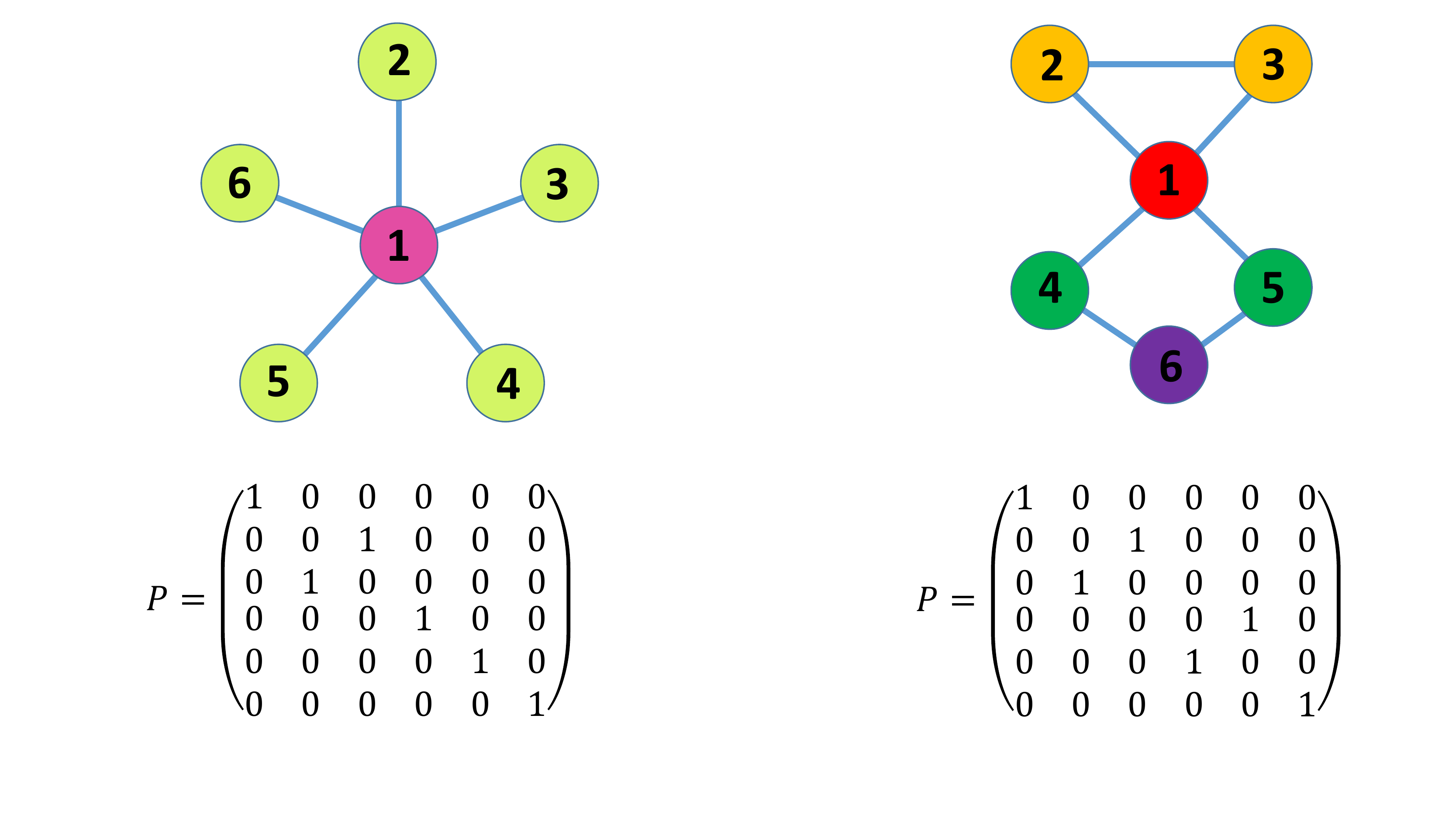}
	\caption{Two examples of networks with $n = 6$ vertices possessing symmetries. The first example, a star, on the left, has the property that any permutation of its leaves is an automorphism. Swapping only leaves $2$ and $3$ has the permutation matrix $P$ described below the graph. For the second example, on the right, swapping vertices $2$ and $3$ and $4$ and $5$ on network preserves the adjacency matrix, with the corresponding permutation matrix $P$ shown on the bottom. The symmetries of the networks reflect on the dynamics in such a way that vertices of the same color have the same dynamics if the natural frequencies satisfy appropriate conditions.}
	\label{sym_examples}
\end{figure}

The automorphisms, in fact permutations $\pi$ of the set $V$ of vertices, may be represented by a permutation matrix $P$, a square binary matrix that has exactly one entry of $1$ in each row and each column and $0$ elsewhere that commutes with the adjacency matrix, $AP = PA$. Consider, for example, Figure \ref{sym_examples}. On the left we have a star consisting of a hub and $5$ leaves. Any permutation of the leaves is an automorphism, in particular, swapping vertices $2$ and $3$ has the matrix $P$ shown bellow the network. The next example, on the right, again with $n = 6$ vertices, is invariant if we swap vertices $2$ and $3$ or if we swap $4$ and $5$. In particular, if we apply both swaps, the corresponding permutation matrix is shown bellow the network in Figure (\ref{sym_examples}).

A permutation can also be used to partition the elements into disjoint sets. Consider the permutation described for the network on the right of Figure (\ref{sym_examples}). If we apply it, it maps $2$ to 3 and 4 into 5. Applying it once again, now 3 goes back to 2 and 5 to 4. Any power of the permutation will swap the vertices in this manner, whereas the remaining vertices will be kept the same. In this way we can partition the vertices into $(2,3)$, $(4,5)$ and the rest of the elements.

Finally, we also need to understand the effect of permuting the Hadamard product of two vectors. Given two $n\times1$ column vectors, $\boldsymbol{v}$ and $\boldsymbol{u}$, the Hadamard product, denoted by $\circ$, is defined as

\begin{equation}
\boldsymbol{v} \circ \boldsymbol{u} =  \dottedcolumn{4}{v_1}{v_n} \circ \dottedcolumn{4}{u_1}{u_n} = \dottedcolumn{4}{v_{1}u_{1}}{v_{n}u_{n}}
\label{hadamard} 
\end{equation}

The product $P \left( \boldsymbol{v} \circ \boldsymbol{u} \right)$, where $P$ is a permutation matrix, arranges all the lines $\left( \boldsymbol{v} \circ \boldsymbol{u} \right)$  into some new order. But this can be obtained by arranging both vector $\boldsymbol{v}$ and $\boldsymbol{u}$ independently but in the same manner (using $P$) and then taking the Hadamard product. Therefore, if we first permute both $\boldsymbol{v}$ and $\boldsymbol{u}$ and then take the Hadamard product, we conclude that $P\left(\boldsymbol{v} \circ \boldsymbol{u} \right) = \left(P\boldsymbol{v} \circ P\boldsymbol{u}\right)$. Note, however, that this distributive property is not shared with more general matrices.

\section{Remote synchronization for phase oscillators}
\label{remote_sync_phase_oscillators}

We will begin by considering a general model of coupled phase oscillators on top of $\mathcal{G}$, given by

\begin{equation}
m\ddot{\varphi}_i + \dot{\varphi}_i = \omega_i + f(\varphi_i) + \lambda g(\varphi_i) \sum_{j=1}^{N} A_{ij} h(\varphi_j),
\label{general_model}
\end{equation}
where $\varphi_i$ and $\omega_i$, for $i=1,2,...,n$, are the phases and natural frequencies of the oscillators, $m$ is a parameter that plays the role of inertia, $\lambda$ is the coupling strength and functions $f$, $g$ and $h$ are the same for all oscillators (later we will give specific examples of these functions for some well know models). Note that this more general model is not covered in \cite{pecora2014}. The trick to accomplish our goal is that equation (\ref{general_model}) can be written in matrix-vector notation as

\begin{equation}
m \ddot{\boldsymbol{\varphi}} + \dot{\boldsymbol{\varphi}} = \boldsymbol{M}(\boldsymbol{\varphi}) = \boldsymbol{\omega} + \boldsymbol{F} +  \lambda \boldsymbol{G} \circ \left(A\boldsymbol{H}\right),
\label{general_model_matrix_form}
\end{equation}
where $\boldsymbol{F}$, $\boldsymbol{G}$ and $\boldsymbol{H}$ are $n\times1$ column vectors with element $i$ given by the corresponding functions calculated at value $\varphi_i$ and $\circ$ denotes the Hadamard product of two matrices, as discussed previously in section (\ref{preliminaries}).

Given an automorphism of $\mathcal{G}$ and its corresponding permutation matrix $P$, it acts on equation (\ref{general_model_matrix_form}) in the following way,

\begin{equation}
mP\ddot{\boldsymbol{\varphi}} + P\dot{\boldsymbol{\varphi}} = P \boldsymbol{M}(\boldsymbol{\varphi}) = P\boldsymbol{\omega} + P \boldsymbol{F} + \lambda P \left[ \boldsymbol{G} \circ \left(A\boldsymbol{H}\right) \right] = P\boldsymbol{\omega} + P\boldsymbol{F} + \lambda \left(P\boldsymbol{G}\right)\circ A\left(P\boldsymbol{H}\right),
\label{perm_model}
\end{equation}
because we have that $P[G\circ (AH)] = (PG) \circ (PAH)$, due to the distributive nature of permutation matrices and Hadamard products, and from the commuting relation between $A$ and $P$, $AP = PA$, we end up with $(PG) \circ (PAH) = (PG) \circ A(PH)$ 

What result (\ref{perm_model}) means is that the action of permuting the vector field using $P$, that represents an automorphism of $\mathcal{G}$, is equivalent to permuting the phases and natural frequencies of the coupled phase oscillators model. If furthermore the natural frequencies are chosen such that $P\boldsymbol{\omega} = \boldsymbol{\omega}$, as happens, for example, when $\boldsymbol{\omega} = b\boldsymbol{k}$, where $\boldsymbol{k}$ is the vector of degrees and $b$ a constant, or trivially if all the oscillators have the same natural frequency, then the vector field (\ref{general_model}) satisfies the relation 

\begin{equation}
mP\ddot{\boldsymbol{\varphi}} + P\dot{\boldsymbol{\varphi}} =\boldsymbol{M}(P\boldsymbol{\varphi}).
\label{perm_phase_final}
\end{equation}

Suppose that system (\ref{general_model})  has a synchronized state, where all the oscillators have the same (time indenpendent) frequency $\dot{\varphi}_i = \Omega$ and $\ddot{\varphi}_i = 0$, implying that $P\dot{\boldsymbol{\varphi}}  = \dot{\boldsymbol{\varphi}}$. This result, together with (\ref{general_model_matrix_form}) and (\ref{perm_phase_final}), lead us to conclude that $\boldsymbol{M}(\boldsymbol{\varphi}) = \boldsymbol{M}(P\boldsymbol{\varphi})$ and therefore $\boldsymbol{\varphi} = P\boldsymbol{\varphi}$ is a solution of this equality.

If $\boldsymbol{\varphi} = P\boldsymbol{\varphi}$ holds true, it also does for any power of $P$. Therefore we can partition the set of vertices by the orbits generated by applying the permutation repeatedly and all of the vertices that belong to a certain partition must have the same phase, which is remote synchronization.

The system of equations (\ref{general_model}) encompass many cases of interest. For $m \ne 0$ it has been used as a toy model to study power grids \cite{filatrella2008,rohden2012,pinto2016b}. When $m = 0$, it encompass models such as the one proposed by Winfree in the late 1960 \cite{winfree1967,winfree1980}, with $f(\varphi) = 0$ and  $g(\varphi)$ and $h(\varphi)$ serving as the phase response curve (PRC) and the influence function, respectively. An ensemble of overdamped Josephson junctions driven by a constant current and coupled through a resistive load \cite{tsang1991} also can be modeled by a system in the form (\ref{general_model}), as well as the recently proposed model \cite{keeffe2016} of coupled oscillatory and excitable elements. It's also possible to make the coupling dependent on the index $i$, such that $\lambda \rightarrow \lambda_i$ (this can be applied, for example, to the model studied in \cite{pazo2011}) or dependent on the index $j$ and placed inside the summation in (\ref{general_model}). For both cases, following the same arguments, remote synchronization can happen if $P\boldsymbol{\lambda} = \boldsymbol{\lambda}$.

Finally, the Sakaguchi-Kuramoto model (\ref{sakaguchi_kuramoto model}) can also be written in the form (\ref{general_model}), by expanding the sine of the phase difference. In this case we get two copies of the last term in (\ref{general_model_matrix_form}), the first one being $\boldsymbol{G} = \cos(\boldsymbol{\varphi})$ and $\boldsymbol{H} = \cos(\alpha) \sin(\boldsymbol{\varphi}) - \sin(\alpha) \cos(\boldsymbol{\varphi})$ and the second ones $\boldsymbol{G} = -\sin(\boldsymbol{\varphi})$ and $\boldsymbol{H} = \cos(\alpha) \cos(\boldsymbol{\varphi}) + \sin(\alpha) \sin(\boldsymbol{\varphi})$, where $\sin(\boldsymbol{\varphi})$ and $\cos(\boldsymbol{\varphi})$ are vectors with elements being the sine or the cosine of the phases.

The conditions discussed previously encompass also more general states that cannot be exactly described as synchronization, but nevertheless still reflect the graph symmetries. One such example is the state called \emph{oscillation death} that happens for the version of the Winfree model studied in \cite{ariaratnam2001,pazo2014,pinto2016}. In this state the system approaches a fixed point with $\dot{\varphi}_i = 0$ for all oscillators, in conformity with our assumptions, such that the phases are frozen at specific angles satisfying $\boldsymbol{\varphi} = P\boldsymbol{\varphi}$. An interesting fact is that an excellent approximation \cite{pinto2016} for this state is given by $\varphi_i = \gamma \omega_i k_{i}^{-1}$, where $\gamma$ is a constant that is independent of the natural frequencies or the structure of the network. It's clear that in this approximation the phases are invariant under permutations that are graph automorphisms when $P\boldsymbol{\omega} = \boldsymbol{\omega}$. A numerical example of this situation is discussed in section (\ref{examples}).

\section{Remote synchronization for Stuart-Landau oscillators}
\label{remote_sync_sl_oscillators}

The Stuart-Landau model for a general network, is given by the system of differential equations

\begin{equation}
\dot{z}_{i} = \left(\alpha + i\omega_i - |z_{i}|^2\right)z_i + \lambda \sum_{j=1}^{N} A_{ij} (z_j - z_i)
\label{stuart_landau_model}
\end{equation}
and describes systems in the neighborhood of a Hopf bifurcation, where the strength of the attraction to the limit cycle is not necessarily strong, such that the amplitude dynamics is important. In the absence of coupling, when $\lambda = 0$, the equation has an unstable fixed point at $z_i = 0$ and a stable limit cycle with radius $\sqrt{\alpha}$ and natural frequency $\omega_i$. Moreover, $\alpha$ is a measure of the strength of attraction to limit cycle and it can be proven that when $\alpha \rightarrow \infty$ we recover the Kuramoto model \cite{gambuzza2013}.

The idea is actually the same that we used for phase models, we first write equations (\ref{stuart_landau_model}) in matrix-vector notation as

\begin{equation}
\dot{\boldsymbol{z}} = \boldsymbol{M}(\boldsymbol{z}) = \left(\boldsymbol{\alpha} + i\boldsymbol{\omega} - \boldsymbol{z}^{*}\circ\boldsymbol{z}\right) \circ \boldsymbol{z} - \lambda L \boldsymbol{z},
\label{ugly_equation}
\end{equation}
where we have introduced the Laplacian matrix $L = D - A$, where $D$ is a diagonal matrix with entries equal to the degrees of the vertices, $\boldsymbol{\alpha}$ is a vector with all its elements equal to $\alpha$ and $*$ denotes the conjugated value.

Applying the permutation $P$ on both sides of (\ref{ugly_equation}), and noticing that $L$ and $P$ commute, since symmetric vertices have the same degrees and therefore $PD = DP$, and assuming that $P\boldsymbol{\omega} = \boldsymbol{\omega}$, we obtain that 

\begin{equation}
P\dot{\boldsymbol{z}} = P\boldsymbol{M}(\boldsymbol{z}) = \left(\boldsymbol{\alpha} + iP\boldsymbol{\omega} - P\boldsymbol{z}^{*}\circ P \boldsymbol{z}\right) \circ P\boldsymbol{z} - \lambda L P \boldsymbol{z} = \boldsymbol{M}(P \boldsymbol{z})
\label{not_so_ulgy}
\end{equation}

If system (\ref{not_so_ulgy}) has a synchronized state with $\dot{\boldsymbol{z}} = 0$, then it implies that $\boldsymbol{M}(P\boldsymbol{z}) = \boldsymbol{M}(\boldsymbol{z})$ and therefore it has remote synchronization, $P\boldsymbol{z} = \boldsymbol{z}$ and we can use permutations that are graph automorphisms to partition the vertices into disjoint sets by repeatedly applying $P$, such that in each one of these sets, vertices possess the same values of $z$.

\section{Numerical examples}
\label{examples}

\begin{figure}
	\centering
	\includegraphics[width=0.7\linewidth]{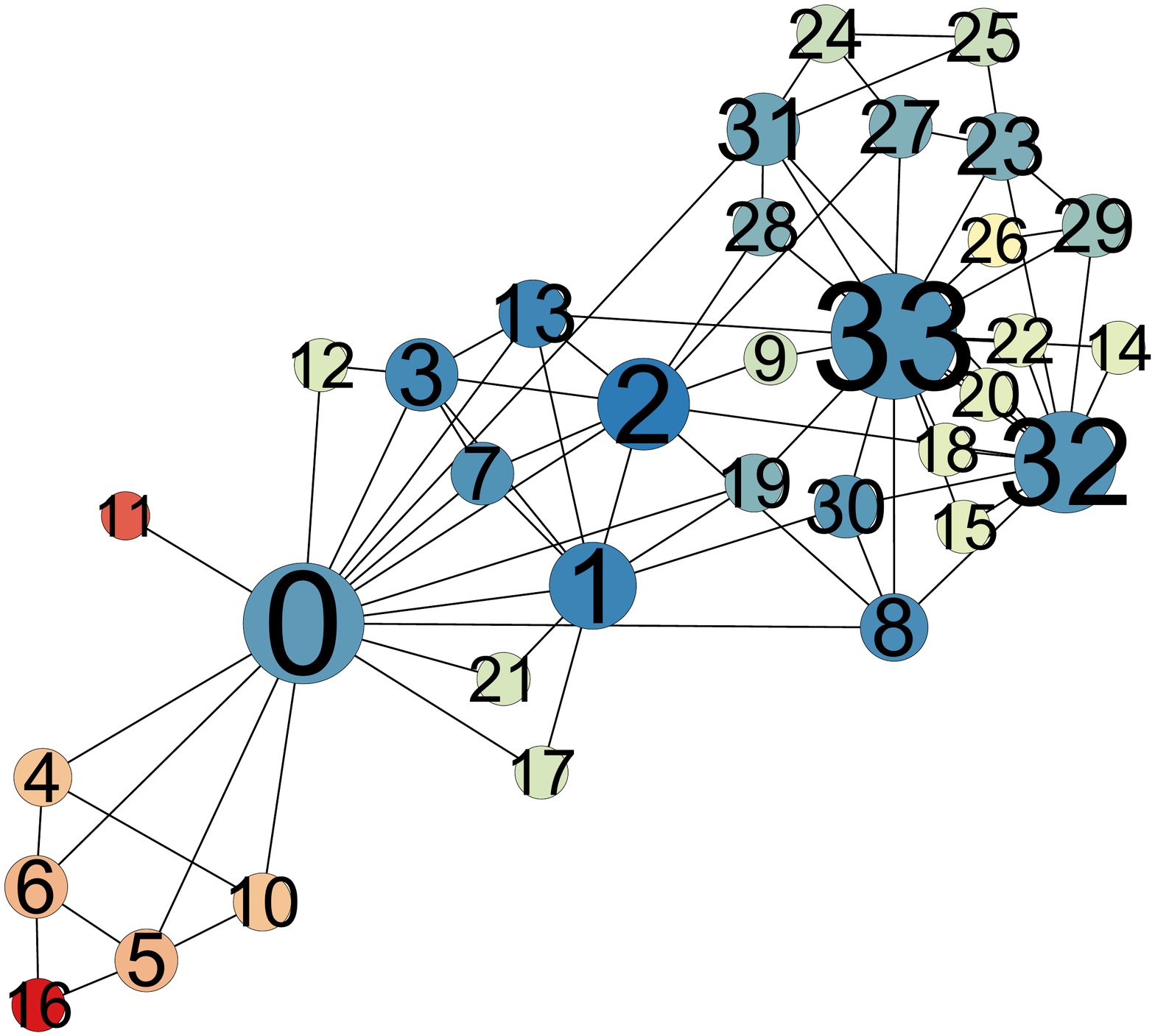}
	\caption{The phases in the Karate network with Kuramoto dynamics and degree-frequency correlation. The vertices size and color are proportional to its degrees and phases $\varphi_i$ (at $t = 100$), respectively. The groups of vertices with the same phase are  $(5,6)$, $(4, 10)$, $(17, 21)$ and $(14, 15, 18, 20, 22)$.}
	\label{remote_sync_kuramoto_karate}
\end{figure}

The purpose of this section is to corroborate numerically what has been discussed so far. As our first example, we employ the same network used in \cite{vlasov2016}, the Karate club network \cite{zachary1977}. It is an undirected network with $n = 34$ vertices and $m = 78$ links. Although for larger networks it starts to be harder and harder to find graph automorphisms, from Figure \ref{remote_sync_kuramoto_karate} we can see the symmetries without too much effort. We have the following sets $(5,6)$, $(4, 10)$, $(17, 21)$ and, somewhat harder to find, $(14, 15, 18, 20, 22)$ such that swapping pairs of vertices that belong to the same set leaves the adjacency matrix invariant and therefore must have the same dynamics.

We simulate the Kuramoto model (equation (\ref{general_model}), with $\alpha = 0$) using $\lambda = 4.0$ and degree-frequency correlation, $\omega_i = k_i$, for the Karate network. At time $t = 100$ we stopped the simulation and painted the vertices in Figure \ref{remote_sync_kuramoto_karate} with colors according to their phases (the size is proportional to the degree). The simulation shows that the vertices with the same dynamics are those predicted by theory. For example, we obtained that $\varphi_{14} = \varphi_{15} = \varphi_{18} = \varphi_{20} = \varphi_{22} = 4.76720476$, $\varphi_{17} = \varphi_{21} = 4.72497040$, $\varphi_{5} = \varphi_{6} = 4.53766005$ and $\varphi_{4} = \varphi_{10} = 4.56300853$.

\begin{figure}
	\centering
	\includegraphics[width=0.7\linewidth]{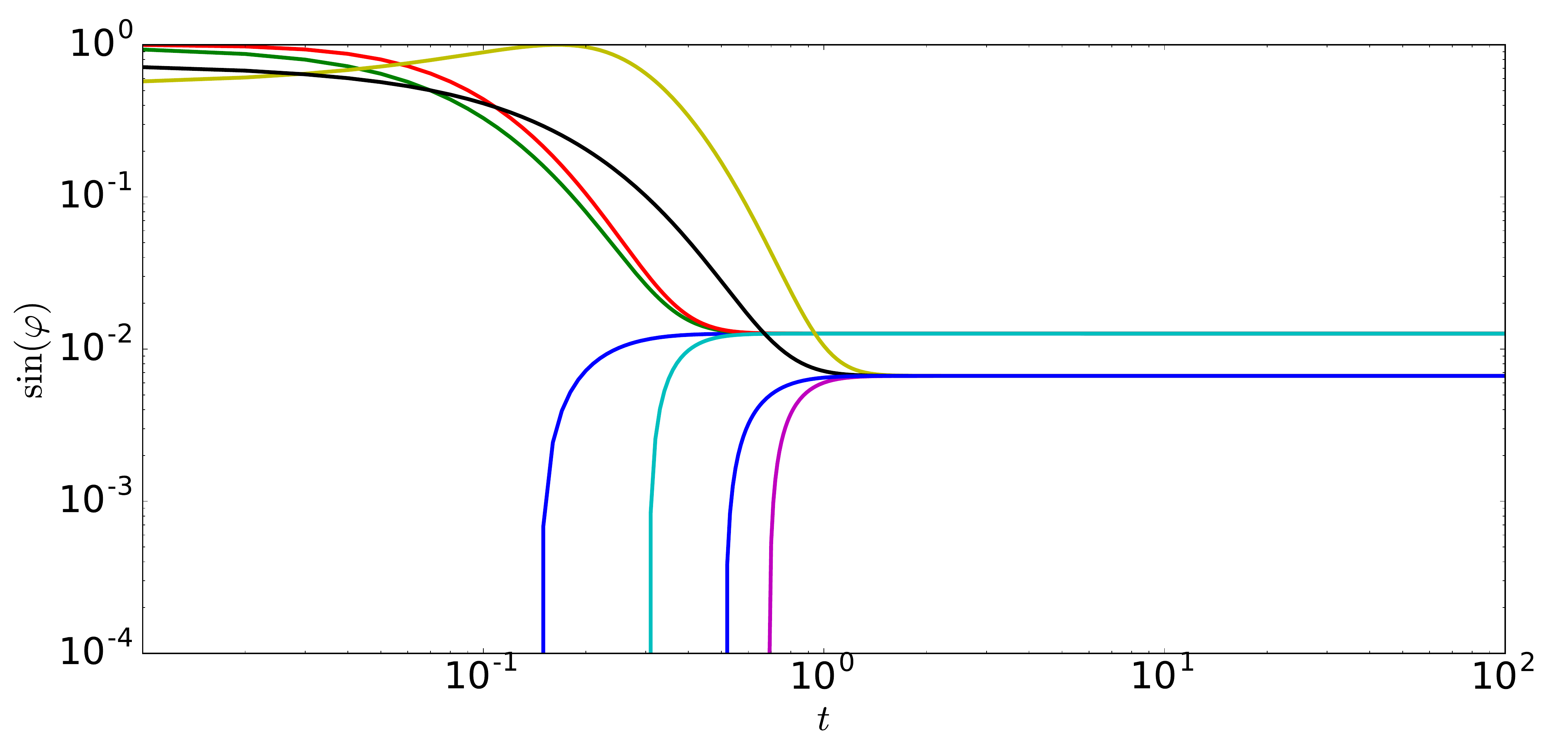}
	\caption{The sine of the phases of two groups, of 4 vertices each, that are symmetric, $(69,70,71,72)$ and $(238, 239, 240, 241)$, in the oscillation death state for the Winfree model \cite{ariaratnam2001,pazo2014,pinto2016}. The natural frequencies were chosen correlated to the squared degrees of the vertices, $\omega_i = k_{i}^{2}$.}
	\label{remote_sync_winfree_macaque}
\end{figure}

Another example is the macaque cortex \cite{harriger2012}, with $n = 242$ vertices and $m = 3054$ edges. For this network, finding by inspection symmetrical vertices is already quite troublesome. We simulate the Winfree model \cite{ariaratnam2001,pazo2014,pinto2016} with parameters chosen specifically to fall in the oscillation death regime. Figure \ref{remote_sync_winfree_macaque} shows the phases for two groups of four vertices each that are symmetric. As expected, each group freezes all of its phases at the same point.

\section{Conclusion}
\label{conc}

We analyzed the issue of remote synchronization, a topic that has attracted a lot of activity in the last few years. Writing the equations of motion in a form that simplifies the analysis of how permutations acts on the vector field, we could find more general models where remote synchronization can be observed, as well as the conditions that the heterogeneous ensembles of oscillators have to satisfy in order to promote RS.

It must be stressed that for general models of phase oscillators, section \ref{remote_sync_phase_oscillators} and for the Stuart-Landau model, section \ref{remote_sync_sl_oscillators}, in the synchronized state the phases (and radius, for SL oscillators) are invariant under the action of permutation matrices $P$ representing graph automorphisms if $P\boldsymbol{\omega} = \boldsymbol{\omega}$ also holds true. This, however, does not mean that all possible solutions display remote synchronization. Consider, for example, the Kuramoto model (equation (\ref{sakaguchi_kuramoto model}) wth $\alpha = 0$) and identical frequencies. In this case the only solution is complete synchronization, with $\varphi_i = \varphi$ for all the oscillators. Obviously this is not remote synchronization, even if it inherited all the symmetries of the graph (a set of identical elements is trivially invariant under any permutation). It's necessary some extra ingredient to avoid that all the phases coalesce such that it can be organized into disjoint sets that necessarily must be invariant under the graph automorphisms. When all the oscillators are identical, this ingredient can be frustration \cite{nicosia2013}. A further option is non-identical natural frequencies, that however satisfies $P\boldsymbol{\omega} = \boldsymbol{\omega}$, as happens when we impose degree-frequency correlation \cite{vlasov2016,bergner2012}. This connect the previous results in the literature with what was proved in here.

The general system (\ref{general_model}), or even the SL model (\ref{stuart_landau_model}), do not exhausts all the possible models where RS can be observed, but the outline proposed here can be used quickly to determine if more elaborated models possess the ability to develop RS. 

Further questions that must be answered include the stability of these symmetrical states due to perturbations, such as vertices or links removal and natural frequencies not exactly invariant under the permutation. Another point is that in large complex networks it is harder to find graph automorphisms. Even for the small Karate club network it takes some time to spot the symmetries mentioned out in the text . Maybe RS can be a tool to find automorphisms. 

Finally, is remote synchronization seem in nature? Many dynamical systems have equations of motion resembling the ones used here, which may indicated the \emph{remote dynamics} is expected to be a common phenomenon.

\section*{Acknowledgements}

The author thank CNPq for the financial support, as well as Edmilson Roque and Thomas Peron for useful discussions. Our numerical computations were done by using the packages SciPy \cite{jones2001} and NetworkX \cite{hagberg2008} for python together with the software Gephi \cite{gephi} for network visualization.

\end{document}